\documentclass[12pt]{report}

\usepackage{graphicx}
\usepackage{float}

\usepackage{amssymb}
\usepackage{multirow}
\usepackage{multicol}
\usepackage[square,numbers]{natbib}
\usepackage{longtable}
\usepackage{rotating}
\usepackage{fancyhdr}
\usepackage{tikz}
\usepackage[tikz]{mdframed}
\usepackage{pdfpages}

\usepackage{color, soul, colortbl}

\usepackage{titlesec} 
\usepackage[left=4cm,right=4cm,top=2.5cm,bottom=4cm]{geometry}

\usepackage{tabularx}

\usepackage{hyperref}

\definecolor{LightGray}{gray}{.85}

\definecolor{DarkRed}{RGB}{163,22,31}
\definecolor{LightRed}{RGB}{247,14,45}
\definecolor{PaleRed}{RGB}{239,134,148}
\definecolor{cpiGray}{RGB}{106,100,100}

\titleformat*{\section}{\color{DarkRed}\normalfont\bfseries\LARGE}
\titleformat*{\subsection}{\color{LightRed}\normalfont\bfseries\LARGE}
\titleformat*{\subsubsection}{\color{PaleRed}\normalfont\bfseries\large}

\def\title#1{\gdef\@title{#1}\gdef\reporttitle{#1}}

\title{Understanding motivations and characteristics of financially-motivated cybercriminals}
\author{Awais Rashid, Claudia Peersman, Chao Chen, Ziauddin Ursani, Joseph Hallett, Matthew Edwards \& Louise Evans}

\renewcommand{\maketitle}{\newpage
\newgeometry{margin = 0in}
\includegraphics[width=3.09in]{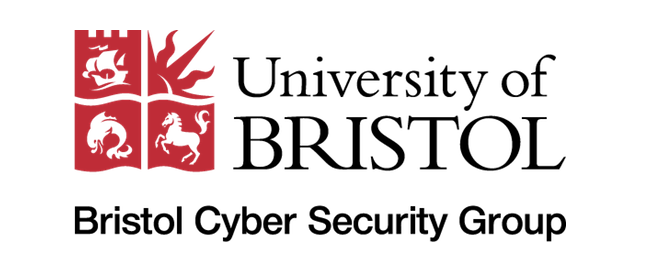}
\setlength{\fboxsep}{0pt}
\hfill \colorbox{cpiGray}{\makebox[3.22in][r]{\shortstack[r]{\vspace{2.75in}}}}%
\vspace{-0.25pt}
\setlength{\fboxsep}{10pt}
\setlength{\fboxrule}{0pt}
\colorbox{DarkRed}{\makebox[8.25in][l]{\hfill \shortstack[r]{\fontsize{16}{16}\rmfamily\color{white} \reporttitle\\%
\fontsize{14}{14}\rmfamily\color{white}}}}%
\setlength{\fboxsep}{0pt}
\vspace{-8.5pt}
\hfill \colorbox{cpiGray}{\hspace{.25in} \parbox{2.97in}{\vspace{3.6in} \color{white} \textbf{Claudia Peersman \\ Emma Williams \\ Matthew Edwards \\ Awais Rashid\\\\   \today \vspace{1in}}
\vfill}}
\let\Title\title
\restoregeometry}

\mdfdefinestyle{SummaryBox}{%
    linecolor=DarkRed,
    outerlinewidth=3pt,
    roundcorner=20pt,
    innertopmargin=\baselineskip,
    innerbottommargin=\baselineskip,
    innerrightmargin=20pt,
    innerleftmargin=20pt,
    frametitlerule=true,
    frametitlerulecolor=DarkRed,
    frametitlefontcolor=white,
    frametitlebackgroundcolor=DarkRed,
    fontcolor=DarkRed,
    frametitle=Summary,
    backgroundcolor=gray!50!white,
    leftmargin=18pt,
    rightmargin=18pt}

\begin{document}

\begin{titlepage}
\maketitle
\end{titlepage}
\newpage

\section*{Understanding motivations and characteristics of financially-motivated cybercriminals:}\subsection*{Revisiting theoretical approaches through the lens of contemporary practitioner experiences}

\textit{\textbf{Background:}} Cyber offences, such as hacking, malware creation and distribution, and online fraud, present a substantial threat to organizations attempting to safeguard their data and information. By understanding the evolving characteristics and motivations of individuals involved in these activities, and the threats that they may pose, cyber security practitioners will be better placed to understand and assess current threats to their systems and the range of socio-technical mitigations that may best reduce these.
\textit{\textbf{Aim:}} The reported work-in-progress aims to explore the extent to which findings from prior academic literature regarding the characteristics and motivations of offenders engaging in financially-motivated, cyber-dependent crime are supported by the contemporary experiences and perspectives of practitioners currently working in the cyber crime field.  
\textit{\textbf{Method:}} A targeted, online survey was developed consisting of both closed and open-ended questions relating to current cyber threats and the characteristics and motivations of offenders engaged in these activities. Sixteen practitioners working in law enforcement-related domains in the cyber crime field completed the survey, providing a combination of qualitative and quantitative data for analysis.
\textit{\textbf{Results:}} Findings demonstrated support for three characteristics previously highlighted in prior research (younger age, being male, and greater computer use) and multiple motivations, with financial gain and a feeling of power and status being the most common. Several nuances in characteristics and motivations were also considered by practitioners related to the particular type of cybercrime engaged in, and how these various aspects appear to be evolving due to the increasing commercialization of ready-to-use toolkits, the growth of cyber-crime as a service, and the involvement of organized crime groups and others in malevolent cyber activity.
\textit{\textbf{Conclusions:}} The reported work provides key insight into the views, perspectives and experiences of current practitioners in the cyber crime field, overlaying theoretical concepts and findings from prior academic literature with elicited perspectives from law enforcement practitioners. As a work-in-progress, it is intended that further work will be undertaken to consider additional data sources that may further validate these findings.

\newpage

\section*{Background}
Cyber offences such as hacking, malware creation and distribution, and online fraud present a substantial threat to organizations and individuals attempting to safeguard their data and information. Understanding the evolving characteristics and motivations of individuals involved in these activities can help practitioners attempting to disrupt or counter cybercriminal activity. Although there is a reasonable body of work that investigates the various motivations and characteristics of cybercriminal offenders, the constantly evolving nature of cyber threats and those involved means that it is vital to revisit prior work in relation to whether previous findings accurately reflect the current experiences of contemporary practitioners working in the field.

This work in progress, therefore, aims to contribute to current understanding of the motivations and characteristics of those involved in cybercriminal activities and the threats that they may pose from a cyber security perspective in order to inform the future development of socio-technical mitigations that can address these evolving threats. Specifically, this paper overviews the findings of prior research regarding the characteristics and motivations of those involved in cybercriminal activities and explores the extent to which these findings are reflected in the current perspectives and experiences of practitioners working in the cyber crime domain. As a work in progress, it is intended that additional data sources will be considered in future work (e.g., the analysis of current case data or data gathered from relevant darknet market (DNM) communities) to examine the extent to which these motivations and characteristics may be reflected in other datasets.

\section*{Related Literature and Theoretical Approaches}
Previous work has investigated the role of a number of characteristics of cyber offenders, including age, gender, Internet usage, and degree of self-control. Although there is some variation in findings, overall research suggests that younger adults (e.g., those in their late teens and twenties) and males are more likely to be involved in cyber criminal activities (e.g., \cite{hyslip2018defining, woo2003hacker, liao2017computer, voiskounsky2003flow, seigfried2015assessing, payne2018using, mcbrayer2014exploiting, moon2010general}), with some suggestion that such individuals also demonstrate higher education levels \cite{hyslip2018defining, woo2003hacker} and spend more time online (e.g., \cite{moon2010general, donner2014low, pyrooz2013gangs, donner2016gender}). 

The various factors that may motivate individuals to engage in cyber criminal activities have also been explored, with motivations ranging from financial reward \cite{hyslip2018defining, woo2003hacker, mcbrayer2014exploiting, odinot2016cybercrime, kerstens2016victim, holt2007subcultural}, curiosity (e.g., \cite{woo2003hacker, voiskounsky2003flow, xu2013computer}), and entertainment \cite{voiskounsky2003flow, odinot2016cybercrime, kerstens2016victim,xu2013computer, ho2012influences} to addiction \cite{woo2003hacker, voiskounsky2003flow, mcbrayer2014exploiting, kerstens2016victim, kranenbarg2017offending}, individual vengeance \cite{hyslip2018defining, voiskounsky2003flow, mcbrayer2014exploiting, odinot2016cybercrime, kerstens2016victim, holt2007subcultural, xu2013computer, ho2012influences, kranenbarg2018cyber}, the pursuit of power or status \cite{woo2003hacker, mcbrayer2014exploiting, holt2007subcultural,xu2013computer, marcum2014hacking}, ideology \cite{woo2003hacker, holt2009attack} and learning, immersion and intellectual challenge \cite{woo2003hacker, voiskounsky2003flow}. 

Unfortunately, research in this area is often constrained by methodological limitations, predominantly due to difficulties in accessing offender populations and other forms of data and a heavy reliance on self-report measures and self-selected samples, which continues to be an issue in many human-focused research fields (see \cite{payne2020disciplinary} for a review). Quantitative data is typically analyzed using regression approaches to determine which factors may predict involvement in cybercriminal activities (e.g., \cite{donner2014low, kerstens2016victim, marcum2014hacking,holt2017exploring}). While this analytic approach is sound, online surveys often utilize student samples (e.g., \cite{donner2016gender}) or focus on younger demographics (see \cite{woo2003hacker} for a review), which can present limitations with regards to the generalizability of findings across wider population groups.

More recently, work has scraped forums or utilized readily available datasets (e.g., DNM Archives, see \cite{dnmArchives}) alongside deploying big data analytics (e.g., data mining, machine learning) to provide a mechanism to continue to improve and develop current understanding of cybercriminal behavior (e.g., \cite{decaryhetu2013reputation}). However, papers using scraped data commonly focus on \textit{market dynamics} and the impact of law enforcement interventions rather than offender motivations or characteristics. In order to minimize the various limitations of these differing approaches, the triangulation of multiple data sources and integration of theoretically-based predictions represents the optimal, albeit more resource intensive, approach. 

A range of theoretical approaches from psychology, criminology and sociology have been applied to understand the factors that underpin the behavior, motivations and characteristics of cyber criminals, with a particular focus on Neutralization Theory \cite{sykes1957techniques}, Self-Control Theory \cite{muraven2006self}, Social-Cognitive Theory \cite{bandura2001social}, Routine Activity Theory \cite{cohen1979social}, and Theory of Planned Behavior \cite{ajzen1991theory}.

At the individual level, Neutralization Theory (NT; \cite{sykes1957techniques}) focuses on how individuals perceive and consider the potential impact of their criminal activities, and the way in which they may distort these perceptions in order to make their activities more morally acceptable. For instance, if individuals believe that engaging in hacking is a morally unethical behavior, but they engage in this activity, then in order to reduce the resultant conflict between their moral perceptions and their behaviors, they can either stop engaging in the behavior altogether (i.e., no longer engage in hacking activities), or they can adjust their attitudes and thoughts to correspond with their behavior (e.g., considering that hacking is not hurting anyone directly and so it must be okay, thus making it a morally acceptable activity). Neutralization theory has been applied to hacking (e.g., \cite{morris2011computer}) and digital piracy activities (e.g., \cite{morris2009neutralizing}), with five primary mechanisms highlighted by which this may occur: (1) denial of responsibility (e.g., 
``I had no other choice''), (2) denial of injury (e.g., ``it doesn't hurt anyone''), (3) denial of victim (e.g., ``the victim deserves it in some way''), (4) condemnation of the condemners (e.g., ``victims are hypocrites''), and (5) appeal to higher loyalties (e.g., the behavior was warranted to achieve a higher purpose).

Self-Control Theory of Crime (SCTC) also considers the influence of individual characteristics, considering that individuals often engage in criminal activities due to failure in their ability to effectively control their behavior \cite{muraven2006self}. Self-control is considered a trait psychological characteristic linked to the ability to avoid and resist temptation (see \cite{ent2015trait}). Individuals low in self-control are considered to be more impulsive, take greater risks, and fail to adequately consider the long-term consequences of their actions. This trait has been linked to cybercriminal activities, such as downloading illegal content \cite{moon2010general, donner2014low, higgins2007digital, nodeland2020impact}.

Other approaches, such as Social Cognitive Theory (SCT; \cite{bandura2001social}) focus on the influence of the wider environment on engagement in cybercriminal activities. For instance, Social Cognitive Theory suggests that individuals learn how to behave in a particular environment based on observing the actions of those around them and the resultant consequences and outcomes of the behaviors that they observe. It is proposed that this social learning process occurs through engagement with particular role models within an individual’s environment, such as observing the behavior of family members. It can also occur via exposure to mass-media content or particular social norms and activities linked to specific geographical locations or communities. This process may not only influence what types of behavior an individual considers to be socially acceptable, but can also lead to direct learning and skill acquisition related to the actual ability to engage in particular types of online activities. For instance, attitudes to digital piracy have been linked to association with deviant peers and the perception that important others wanted or expected the individual to engage in these activities \cite{d2005music, kirwan2013cybercrime, malin2009adolescent}. Social-ecological approaches to deviant online activities, such as cyber bullying, also highlight the potential role of these various social factors in influencing individual behavior \cite{cross2015social}.

Routine Activities Theory (RAT; \cite{cohen1979social}) similarly considers the wider environment in influencing engagement in cybercriminal activities, referring to three main elements that must align in order for a crime to occur: (1) the existence of an attractive target (e.g., credit card details stored in a particular location) (2) the presence of a motivated offender (e.g., an individual who becomes aware of this data or actively seeks it out), and (3) the lack of a capable guardian (e.g., a lack of sufficient technical protections). In this way, the existence of a sufficiently attractive target and the use of poor protective mechanisms can, in turn, influence the behavior, characteristics and potential motivations of an offender to commit a particular crime (e.g., a particular activity may represent an intellectual challenge, or poor protections may mean that complex technical skills are not necessary).

Finally, the Theory of Planned Behavior (TPB; \cite{ajzen1991theory}) overlaps with many of the aspects discussed above, considering three primary mechanisms through which an individual's intentions to engage in a particular behavior may be influenced: (1) individuals attitudes towards a behavior, which represents how positively or negatively they evaluate the behavior and those who engage in it, (2) perceived social norms related to engaging in that behavior, and (3) the perceived ease of engaging in the behavior (termed perceived behavioral control). Overall, the use of the TPB as a framework to understand and change behavior has been supported in multiple domains, although effects are generally found to be stronger for intentions to engage in particular activities compared to actual behavioral responses (see \cite{sheeran2016intention} for a review). Although the TPB has been predominantly considered in relation to what factors influence engagement in secure online behavior, there has also been a very limited consideration of how these concepts may apply to cybercriminal behaviors, such as hacking (e.g., \cite{rennie2007advanced}). Previous research has also shown that the perceived likelihood of getting caught when engaging in a cybercriminal activity can influence an individual's attitudes. For instance, Nandedkar and Midha \cite{nandedkar2012won} assessed the presence of optimism bias in relation to music piracy and suggested that greater optimism with regards to potential risks (i.e., an individual thinking that they were less likely to get caught compared to other people) influenced resultant attitudes towards digital piracy. 

\section*{Research Aims}
The current study reports work in progress to examine the contemporary characteristics and motivations of individuals engaged in cyber criminal activities. In particular, it aims to explore the extent to which findings of prior research are supported by the experiences and perspectives of practitioners currently working in the cyber crime field, with a view to understanding current threats and evolving motivations and characteristics of individuals involved in such activities.

A targeted survey approach is used to examine the experiences of individuals working within law enforcement-related contexts (at either regional, national, or international level), offering a unique insight into the practical experiences of individuals working with this offender group. Our approach aims to overlay theoretical concepts and findings from prior academic literature with elicited perspectives from law enforcement practitioners, with future work aiming to also triangulate these findings with data gathered from DNM communities. Linking these aspects provides a novel perspective on cybercriminal motivations and characteristics that will provide a useful framework for developing appropriate socio-technical interventions (e.g., diverting early stage offenders to more positive outlets for their skills). The characteristics and motivations for cyber offenders identified by prior literature and theoretical approaches are shown in Table \ref{tab:prior} and provide a basis for comparison with the views and experiences of current practitioners working in the cyber crime field. This allows us to explore the following research question:

\textit{Research Question: To what extent do the perspectives of practitioners regarding the characteristics and motivations of cyber offenders reflect themes identified in current research literature?}

\begin{sidewaystable}
\centering
\caption{Characteristics and motivations of cyber offenders based on findings of prior literature and primary theoretical approaches }
\scriptsize
\begin{tabular}{ p{20mm}p{30mm}p{50mm}p{50mm}}
\hline
Category One & Definition & Category Two & Themes\\
\hline
\multirow{2}{*}{Characteristics} & A feature or quality belonging typically to an individual & Demographic characteristics: aspects relating to the structure of a population and the different groups that make them up & Age, gender, education, internet/computer use \\ 
& \\
& & Psychological characteristics: psychological factors or traits that may influence behavior, such as personality, attitudes, beliefs & Neutralization of activity (denial of responsibility, denial of injury, denial of victim, condemnation of condemners, appeal to higher loyalties) (NT), Attitude toward cybercriminal activity (TPB) \\
\hline
\multirow{4}{*}{Motivations} & A reason or reasons for acting or behaving in a particular way & Context: the circumstances that form the settings of an event or activity that may contribute to behavior & Self-control (SCTC), Criminal activity in those around them (SCT) \\
& \\
& & Intrinsic: undertaking a behavior to achieve some form of internal satisfaction or reward, such as enjoyment & Addiction/hedonic gratification, excitement/entertainment, curiosity, learning/immersion, feeling of power/status/ego, ideology, vengeance \\
& \\
& & Extrinsic: undertaking a behavior to achieve some form of external reward, such as money & Financial gain, peer recognition* \\
& \\
& & Facilitators: Facilitating conditions that may contribute to motivations & Lack of a capable guardian preventing activity (RAT), Presence of a desirable target (RAT), Ease of engaging in activity (TPB), Likelihood of getting caught (\textit{TPB via influencing attitudes}), Severity of impact of getting caught (\textit{TPB via influencing attitudes}) \\
\hline
\end{tabular}
\label{tab:prior}
\end{sidewaystable}

\newpage 

\section*{Method}
\subsection*{Survey Design}

A targeted online survey was developed using the Qualtrics online survey platform (\url{qualtrics.com}). The survey included a combination of closed and open-ended questions and focused on participants' views and experiences related to the following aspects:

\begin{enumerate}
    \item \textbf{Current cyber threats:} What participants considered the three most significant financially-motivated cyber-crime threats to be (two open-response questions).
    \item \textbf{Offender characteristics:} Whether participants felt that there were particular characteristics of cybercriminal offenders and what these were (\textit{one closed question based on specific characteristics identified in prior literature, with an open-response option for elaboration and additional characteristics}), whether they considered these characteristics to be changing, and whether they considered these characteristics to differ according to different cyber crime threats (\textit{two open-response questions}).
    \item \textbf{Offender motivations:} The extent to which participants considered different factors to represent a motivation for becoming involved, or continuing involvement, in cybercriminal activities (\textit{two closed questions based on specific motivations identified in prior literature with each motivation listed responded to on a scale of 1 (never a motivation) - 5 (almost always a motivation) and an open-response option for elaboration and additional characteristics}), whether they considered these motivations to be changing, and whether they considered these motivations to differ according to different cybercrime threats (\textit{two open-response questions}).
\end{enumerate}

The survey was circulated via email to relevant public and private sector bodies focused on investigating cybercriminal activity and relevant practitioner networks within the law enforcement community. This provided potential respondents with information about the study, the contact details of the researchers, and a weblink for those interested in participating. On accessing the survey, potential respondents were presented with further detail on the research. It was made clear that the survey was voluntary and that participants were under no obligation to complete it. The survey was also anonymous, with no identifying information collected and the data was accessed and analyzed solely by members of the research team at the [removed for peer review]. Participants provided informed consent by indicating that they were happy to proceed before accessing the main survey. The research was granted ethical approval by the relevant institutional review board. A copy of the survey is shown in the Appendix. The qualitative and quantitative data was analyzed in relation to the primary offender motivations and characteristics displayed in Table \ref{tab:prior} to identify the extent to which it supported these various concepts.

\subsection{Participants}
Sixteen participants completed the survey between November 2019 and January 2020. Of these, fourteen stated that they worked in the public sector, one in the private sector, and one in both. Eleven participants worked in investigation roles, four in intelligence roles and one in a management role, with the majority (11 participants: 69\%) working at a regional level. The remainder worked at national or inter-national levels. The majority had worked in their role for at least one year (10; 63\%). Characteristics are also shown in tabular form in the Appendix. 

\section*{Results}
\subsection*{What is the threat?}
Participants were asked to list what they considered to be the top three most significant financially motivated cybercrime threats in order of seriousness (loss or damage). Overall, ransomware was considered to be the most significant threat, mentioned by eleven participants (69\%; with 8 specifying it as their top threat). However, although this provides an indication of how many participants discussed this threat, the low sample size prevents any conclusions being drawn based on quantitative data. Multiple other threats were also mentioned and are shown in Table \ref{tab:threats}. Interestingly, when describing their reasoning for the identification and ranking of these threats, participants highlighted the differing impact of offender characteristics on such threats (e.g., the degree of technical knowledge required), as well as intersections with factors likely to influence motivations, such as the potential low likelihood of being caught and the relative ease with which a crime can be committed. For instance, participant 2 stated \textit{``phishing and compromise of accounts... This is a high volume, low-tech criminal strategy that reaps high returns at low risk. In a month I would see multiple reports of this activity, and maybe one or fewer instances of ransomware/Trojan. I rank Ransomware second as it requires greater tech knowledge and is harder to deploy. However, it is likely to be significantly more damaging in a single incident. I rank Trojans as third threat as they require greater tech knowledge than the other two, and I do not believe them to be as common.''} Similarly, in relation to carding, participant 6 highlighted that \textit{``suspects largely seem to escape any law enforcement intervention. Massive financial loss. Details are available to buy with a little dark web knowledge''}. Finally, participant 12 discussed the necessity to understand the interactions between these activities: \textit{``Banks repeatedly list BEC as major crime in terms of financial loss. However, the data needed for the groundwork of these crimes are (I believe) built on data and access gained through malware infections. Ransomware is the fall-back position to make money from systems that have been compromised via malware.''}

\begin{sidewaystable}
\centering
\caption{Top three most significant financially motivated cyber crime threats reported by participants (\#Resp.: Number of respondents).}
\scriptsize
\begin{tabular}{cccccc}
\hline
Threat One & \#Resp. & Threat Two & \#Resp. & Threat Three & \#Resp. \\
\hline
Ransomware & 8 & Ransomware & 2 & Cryptocurrency vulnerabilities & 2 \\
Business email compromise & 2 & Network intrusion & 2 & Trojans & 1 \\
Carding & 1 & Drugs & 1 & Leaked credentials & 1 \\
Drugs & 1 & Carding & 1 & Phishing & 1 \\
Fraud & 1 & DDoS for hire & 2 & Carding & 1 \\
Phishing & 1 & Phishing & 2 & Hacking & 1 \\
Pharming & 1 & Personal data hacks & 1 & Exploits & 1 \\
- & - & Compromised credentials & 2 & DDoS & 2 \\
- & - & Malware & 1 & Selling data & 1 \\
- & - & Illegal marketplaces & 1 & Ransomware & 1 \\
- & - & - & - & Mandate fraud & 1 \\
- & - & - & - & Remote access trojan & 1 \\
- & - & - & - & Organized crime proliferating malware & 1 \\
\hline
\end{tabular}
\label{tab:threats}
\end{sidewaystable}

\subsection*{Cyber offender characteristics}
Respondents were asked, based on their experience, whether they felt there were particular characteristics of offenders involved in cybercriminal activities. Table \ref{tab:char} shows the cyber offender characteristics that were referenced by at least 50\% of respondents, namely age, gender and extent of computer use. The remaining characteristics listed in the survey (e.g., education level, employment, engagement in offline crime, and personality) did not meet this threshold and so were not supported by the majority of respondents.

\begin{table}
\caption{Cyber offender characteristics referenced by at least 50\% of respondents (\#Resp.: Number of respondents).}
\centering
\scriptsize
\begin{tabular}{lcl}
\hline
Characteristics & \#Resp. & Example descriptive elaboration provided by respondents \\
\hline
Gender & 14 (87\%) & ``Majority male'', ``never dealt with a female subject''\\
Extent of computer use & 12 (75\%) & ``High level of use'', ``always connected'', ``self-built'', ``self-taught''\\
Age & 11 (69\%) & ``Younger age'', ``under 35'', ``easily manipulated due to young age'' \\
\hline
\end{tabular}
\label{tab:char}
\end{table}

When reasoning about their choices, participants highlighted the various nuances, such as crime type, that may influence these characteristics. For example, participant 14 stated ``\textit{characteristics differ depending on the type of cybercrime. Cybercrime resulting in inconvenience can often be younger, more local offenders. Strongly acquisitively driven cybercrime is often more organized, typically originating from overseas… ransomware appears to be more technically capable offenders directly involved (many are penetrative, manual attacks rather than automated malware deployed by unskilled offenders). Mandate fraud is less technically demanding; are they simply buying compromise lists, or employing (contracting to?) a cybercriminal to perform the compromise}'', whilst participant 16 stated 
``\textit{I would suggest Ransomware is often part of an organized crime group - but I have no experience of their characteristics. DDoS and Remote Access Trojans are far more accessible, easy to setup and to deploy and more likely to be the type of offender characteristics I've mentioned.}'' Similarly, participant 5 stated ``\textit{the financially motivated are not typical of the above and quite often involved in organized criminality}'' and participant 2 highlighted that the ``\textit{offender profile above is more likely to be involved in network intrusion. If they can be recruited/exploited by nation state or OCG they may be involved in ransomware/Trojan. I suspect that those involved in business email compromise/phishing are not particularly technically proficient.}'' This view was echoed by participant 11, who suggested ``\textit{maybe slightly older teenagers are manipulated into Serious Organized Crime but they all begin at an early age}'' and by participant 12: ``\textit{BEC doesn't require high end computer programming \& techical understanding (albeit highly skilled fraudsters once they have a person `on the hook')… Malware writes are mostly now coders for hire --- good skills, poor morality}''.

Other participants did not consider there to be set characteristics for such offenders, such as participant 3 who highlighted ``\textit{I believe anybody is capable of committing cybercrime, there are no `set' characteristics associated}'', and participant 7, who considered that any such characteristics are likely to diminish in the future: ``\textit{the opportunity for all persons (current non cyber criminals) to learn how to commit cyber crime is prevalent and available. The attraction of anonymity surrounding access to the dark web and crypto currency makes committing this type of criminality more appealing. Therefore opening the door to creating an alternate cyber criminal profile.}'' Participant 6 echoed this view: ``\textit{I think as cyber-crime becomes more widespread and knowledge improves on a large scale then there is more chance of different character types becoming involved}''.

\subsection*{Cyber offender motivations}
Respondents were also asked, based on their experience, whether they felt there were particular motivations for offenders involved in cybercriminal activities. Table \ref{tab:mot} shows the motivations that were referenced as sometimes, often or always a motivation by at least 50\% of respondents.

\begin{table}[h!]
\caption{Cyber offender motivations referenced by at least 50\% of respondents (\#Resp.: Number of respondents).}
\centering
\scriptsize
\begin{tabular}{lc}
\hline
Motivation & \#Resp. \\
\hline
Feeling of power/status/ego	& 16 (100\%) \\
Financial gain &	16 (100\%)\\
Learning/immersion&	13 (81\%)\\
Perceived acceptability of cyber crime (i.e., ``there are no real victims'')&	13 (81\%)\\
Vengeance&	13 (81\%)\\
Curiosity&	12 (75\%)\\
Exposure to cyber criminal activities \& people engaged in them	&11 (69\%)\\
Perceived likelihood of getting caught&	10 (63\%)\\
Addiction or other hedonic gratification&	10 (63\%)\\
\hline
\end{tabular}
\label{tab:mot}
\end{table}

Respondents expanded on these motivations in their open-ended responses, again highlighting the role of different cybercrime types in influencing these motivations, and the inherent relationship between potential characteristics and motivations. For example, participant 15 stated ``\textit{with DDoS attacks, it can often be jointly motivated by money and knowing that they have the ability to disrupt and cause damage. For pharma/phishing it's mainly financially motivated… I think money will always drive it, with status/recognition following closely behind. For younger offenders it seems that developing skills/reputation is the driving force.}'' Similarly, participant 6 considered ``\textit{it depends on the crime type. There is a correlation between those with fantastic computer knowledge being introverted and often on the autistic spectrum. Their motivations might not necessarily be purely financial and the money could almost be a by-product of their success. Adam Mudd would be a good example, sold DDOS services but his main motivation seemed to be enjoying the success of being really good at something.}''

The growth in so-called cybercrime as a service was also considered to reflect current and future motivations, with participant 1 highlighting ``\textit{cybercrime is becoming more of a service with the main motive being financial and better organized across international OCG's}'' and participant 6 stating ``\textit{the wider range of people who become involved could mean that wider motivations are changing}''. Whilst some participants considered that the focus of cybercrime is ``\textit{becoming more financially motivated}''(participant 1), others considered that such motivations may continue to change based on perceived outcomes, such as participant 16 who suggested that 
``\textit{some motivations might be reinforced and become more significant. E.g. the more victims who pay ransomware demands will obviously provide the offender financial reward and they are likely to continue or escalate (the same is true of any crime with financial reward)}'', as well as wider motivations linked to hostile state actors ``\textit{there may be a secondary motivation to cause disruption on behalf of a nation state, rather than being the sole financial benefactor of the attack}''(participant 2). Finally, the role of online communities was also highlighted by participant 3, who stated 
``\textit{as online communities grow (especially anonymous ones, Tor and other darkwebs) people feel the need to impress these new communities}''. \newpage

\section*{Conclusions}
This work in progress explores the characteristics and motivations of offenders involved in cybercriminal activities. A number of characteristics and motivations were identified based on previous literature and the convergence of these with the current experiences of law enforcement practitioners was examined using a targeted online survey. Overall, findings demonstrated support for three characteristics already highlighted in prior research (younger age, being male, and greater computer use) and multiple motivations, with financial gain and a feeling of power and status being the most common motivations. Several nuances in characteristics and motivations were also considered by practitioners related to the particular type of cybercrime engaged in, and how these various aspects appear to be evolving due to the increasing availability of tools, the growth of cyber-crime as a service, and the involvement of organized crime groups and nation states in malevolent cyber activity.  

Such findings are not only relevant to the academic research community, but also those working as practitioners within the cyber security field. In particular, understanding how characteristics and motivations may be evolving in the current threat environment, how they relate to different cyber threats, and the potential impact that this may have in relation to the optimal protection of systems and data. For example, the perceived prevalence of ransomware, the financial motivations linked to this, and the potential growth of individuals involved in such activity, including the involvement of organized crime groups and nation states and the provision of cybercrime as a service. The continuing professionalization of such activities, and the relative ease of access to technical skills for criminal groups by ‘contracting out’ these aspects, is important to consider when mitigating against these threats. The continued and wide-spread use of so-called ‘low tech’ activities, such as phishing, is also important to consider, particularly when tools to help people with little knowledge ‘enter’ this area are readily available. Indeed, continued development of socio-technical mitigations to phishing and similar threats should remain an active area of research (e.g., \cite{williams2006virtually,ferreira2019persuasion}). The focus on financial gain highlights the primary threats to organizations in this respect, and the potential for some offenders who are initially motivated by intellectual challenge, learning or status to be drawn in to organized crime or other financially-motivated aspects due to their technical skill continues to present an issue. 

In this sense, these findings also provide useful information regarding the continued development of interventions that aim to reduce cyber offending and redirect individuals to positive career paths (e.g., \cite{hmgovernment2016}). These findings highlight the continued perspective that offenders are from younger age groups, are primarily male, and can be vulnerable to exploitation by organized crime groups and others who aim to use their technical skill for financially-motivated (and indeed other forms) of cyber criminal activities. The role of perceptions of power, status and ego as key motivations for cyber offenders are also relevant here, since this may not only influence the likely activities (and potential targets) of cyber offenders, but also the requirement to further understand the communities, relationships and structures that encourage, or indeed discourage, these activities.

\subsection*{Limitations and Next Steps}
It should be acknowledged that this data is based upon the responses of a relatively small number of practitioners working in cybercrime related roles within law enforcement contexts in a particular geographic region. This presents a research limitation, but access to such practitioners is a difficult task and they have provided a diverse range of views and experiences in the qualitative data collected. It should also be acknowledged that responses were self-report. This work focused on analyzing the views, perspectives and experiences of practitioners working in this field and this approach was suitable to achieve this, but it is possible that practitioner views are, to a degree, influenced by wider stereotypes, media information and research findings. It is also likely that practitioner views will be somewhat biased towards those cybercriminal activities (and the individuals engaged in them) that they are aware of (i.e., those individuals who have been caught or who are known to authorities). As such, these findings would benefit from further validation using additional data sources, which continuing work is aiming to address this.

\newpage

\section*{Appendix}
\label{sec:appendix}

\begin{table}[h!]
\caption{Participant characteristics (\#Resp.: Number of respondents).}
\centering
\scriptsize
\begin{tabular}{lc}
\hline
Characteristic & \#Resp. \\
\hline
Public sector & 14 \\
Private sector & 1 \\
Investigation role & 11\\
Intelligence role & 4 \\
Management role & 1 \\
Regional level & 11 \\
National level & 1 \\
International level & 4\\
In role less than a year &6\\
In role 1--2 years & 4 \\
In role 3--5 years & 3\\
In role more than 5 years & 3\\
\hline
\end{tabular}
\label{tab:part}
\end{table}

\textbf{Online survey questions}
\begin{description}
    \item How long have you worked in the cyber crime area? \textit{Less than a year/1-2 years/3-5 years/More than 5 years.} 
    \item Do you work in the public or private sector? \textit{Public/Private/Other.}
    \item What is your main area of focus? \textit{Intelligence/Investigation/Forensics/Other.}
    \item What is the main scope of your work? \textit{International/National/Regional/Other.}
    \item What has been your primary role?
    \item Country?
    \item How do cybercrime offences most frequently come to the attention of your organization? \textit{Through a direct complaint to your organization (e.g., directly from a victim of cyber crime)/Through a centrally-coordinated national reporting mechanism/Through international reporting mechanisms/Unsure/Other.}
    \item In your opinion, what do you consider to be the three most significant financially-motivated cybercrime threats? Please rank them in terms of seriousness (loss or damage).
        \begin{description}
        \item If you can, please elaborate on your views (e.g., why you think that/what your beliefs are based on).
        \end{description}
    \item From your experience, do you feel that there are particular characteristics of offenders involved in cyber-dependent crime? Please tick all that apply and elaborate on your responses if possible: \textit{Age/Gender/Education level/Employment/Engagement in offline crime/ Personality/Extent of computer use/Other.}
        \begin{description}
        \item If you can, please elaborate on your views (e.g., why you think that/what your beliefs are based on).
        \end{description}
    \item From your experience, do you feel that offender characteristics differ across the three cybercrime threats that you identified and if so, in what way?
    \item From your experience, do you feel that offender characteristics for these forms of crime are changing and if so, in what way? 
    \item From your experience, to what extent do you feel that each of the below represent a motivation for \textbf{becoming} involved in cyber-dependent crime? On a scale of 1-5, with 1 being `Never a motivation' and 5 being `Almost always a motivation': \textit{Financial reward/Desire for status, power or recognition/Addiction or other hedonic gratification/Desire for vengeance/Curiosity related to information or technical protections/ Immersion in activity, developing skills and expertise/Previous victim of cyber crime/Exposure to cyber criminal activities and the people engaged in them/Perceived likelihood of getting caught/Perceived acceptability of cyber crime (e.g., ``there are no real victims'')/Other.}
        \begin{description}
        \item Feel free to elaborate/add more options here:
        \end{description}
    \item From your experience, do you feel that offender motivations differ across the three cybercrime threats that you identified earlier?
    \item From your experience, do you feel that offender motivations for engaging in these forms of crime are changing and if so, in what way?
    \item Finally, is there anything else that you would like to add?
\end{description}

%
%
%
 \bibliographystyle{splncs04}
 \bibliography{STAST}

\end{document}